# Evolutionary Model of the Growth and Size of Firms

Joachim Kaldasch

EBC Hochschule Berlin

Alexanderplatz 1, 10178 Berlin, Germany

(Email: joachim.kaldasch@international-business-school.de)

**Abstract**

The key idea of this model is that firms are the result of an evolutionary process. Based on demand and supply considerations the evolutionary model presented here derives explicitly Gibrat's law of proportionate effects as the result of the competition between products. Applying a preferential attachment mechanism for firms the theory allows to establish the size distribution of products and firms. Also established are the growth rate and price distribution of consumer goods. Taking into account the characteristic property of human activities to occur in bursts, the model allows also an explanation of the size-variance relationship of the growth rate distribution of products and firms. Further the product life cycle, the learning (experience) curve and the market size in terms of the mean number of firms that can survive in a market are derived. The model also suggests the existence of an invariant of a market as the ratio of total profit to total revenue. The relationship between a neo-classic and an evolutionary view of a market is discussed. The comparison with empirical investigations suggests that the theory is able to describe the main stylized facts concerning the size and growth of firms.

## 1. Introduction

We want to treat firms as input-output-systems with the aim to create, produce, distribute and finally sell products (and services) in order to make profit. Usually firms consist of several business units responsible for the corresponding products. While firms represent the supply side of a market, the demand side can be characterized by agents (consumers) who are interested in purchasing these products. The key idea of this paper is that the interaction of the supply and demand side in a free market can be considered as an evolutionary process [1].

Evolution is a preferential growth process that consists of three sub-processes: variation, selection and reproduction. The demand side in form of agents willing to purchase a good have the choice between different competing variants (products, brands) of the same good, while we confine here to consumer goods (e.g. shirts, computers, fuel etc). The agents select and purchase products according to their own preferences, usually not exactly known by the supply side. Some products have a higher purchase frequency than others, creating a trend. In order to make profit, firms preferentially reproduce the best-selling products. But firms not just follow the trend; they also use the best-sellers as the basis for new variants of the good. These new variants increase the assortment, while other products may disappear. The cyclic process of selection, reproduction and variation leads after a sufficiently long time to an adaptation of the good such, that the products offered by firms have features wanted by the majority of the agents. So both sides of a free market are interested only in there own selfish goals. But both get what they want: income (profit) and appropriate products. This is what Adam Smith ingeniously realized in the early steps of a free market in England and ascribed this to the action of an invisible hand [2]. The evolutionary approach derived here suggests that the invisible hand is nothing else than evolution.

This statement has considerable consequences. One is that products in a free market play the same role as species in a biological evolution. Form biological evolution it is known that the evolution of the number of individuals of a species (its size) is governed by a replicator dynamics [3]. As will be shown below the dynamics of products is also determined by a replicator dynamics similar to species. It turns out that in this case the size of a business unit grows in proportion to its previous size, while the growth rate is a random variable. This effect is known as the "Law of Proportionate Effect", introduced by Gibrat [4,5] in order to understand the size distribution of firms. The evolutionary model established below derives Gibrat's law and suggests that it is a direct consequence of the evolution of a free market. However, this law applies only to products.

As noticed above, firms consist of a number of business units and have therefore several sources of income. The firm evolution is determined on the one hand by the growth of their products, while each is governed by a multiplicative growth process. But firms are more than just the sum of their products. Firms have the ability to take advantage from their size to create or purchase new products. In biological terms, firms are not "species", they are equivalent to "genera". And from biological genera it is known that their growth is governed by an additionally growth process that is proportional to size of a genus, called preferential attachment [6]. From an evolutionary point of view the size distribution of a firm can therefore be expected to be governed by two growth processes: the proportionate growth of the individual products and a preferential attachment of new products.

Exactly these two ingredients are the basis of the current model of firm growth [7-11]. But these models postulate that the two processes are in action. The main advantage of the evolutionary model is that the growth processes can be traced back to the relation between demand and supply of a free market. It allows establishing approximately the size and growth rate distribution of products and firms. But it also allows a consideration of the price distribution of a good and the number of firms in a market (market size). Taking into account the characteristic property of human activities to occur in bursts, the model try to give an understanding of the size-variance relationship of the growth rate distribution.

The paper is organized as follows. The next section is devoted to a presentation of the evolutionary model. In order to show its applicability a comparison with empirical investigations is performed, followed by a conclusion.

## 2. The Model

We want to consider a consumer goods market. The demand side of a market can be characterized by an ensemble of agents who are interested in purchasing a product, denoted as market potential $M$. Note that goods can be separated into durable goods with a long lifetime (in the order of years) and non-durables (in the order of weeks). In order to derive the size and growth of manufacturing firms we have to study the functionality of a free market.

### 2.1. The Static Market

*The supply side*

The supply side of a market is determined by a number of different variants of a good, denoted here as products (brands), having similar utility properties. They are produced and distributed by manufacturers, while each brand is assigned to a business unit. We want to indicate the products (and the corresponding business units) with index $i$. The total number of different brands in a market is $N$. The absolute number of units of the $i$-th brand sold per unit time is denoted $Y_i$, while $S_i$ indicates the number of supplied units per unit time. In order to establish a model with continuous variables we want to introduce densities, where absolute numbers are scaled by the market potential: $y_i=Y_i/M$ and $s_i=S_i/M$.

The financial value of the $i$-th brand is determined by its nominal price $p_i$. The time dependent mean price is determined by:

$$\langle p(t) \rangle = \frac{1}{y_t(t)} \sum_{i=1}^{N} y_i(t) p_i(t)$$
(2)

while the brackets indicate the average over sold units. The total unit sales and total supply flow can be obtained from:

$$y_t(t) = \sum_{i=1}^{N} y_i(t); \quad s_t(t) = \sum_{i=1}^{N} s_i(t)$$
(3)

Firms may consist of a number of business units. The unit sales of the $j$-th firm scaled by the market potential $x_j$ are given by:

$$x_j(t) = \sum_{i} y_i(t)$$
(4)

where $j$ indicates the firm and the total number of firms is $N_f$.

Business units can be considered as input-output systems. The output is the supply of products and the input is determined by so-called factors (capital and labour). The physical flow is related to a financial flow. The balance of the financial input flow (revenue $R_i=p_iy_i$) and the output flow (costs $C_i$) determines the profit per unit:

$$g_i = p_i - c_i \quad (5)$$

where the costs per unit are $c_i=C_i/s_i$. The costs per unit can be considered to be a function of the supply flow. Since $s_i$ is small, the total costs $C_i$ of a business unit can be expanded as a function of the density of supplied products as:

$$C_i(s_i) \cong c_{i0} + c_{i1}s_i + c_{i2}s_i^2 + ... \quad (6)$$

where the first term represents fixed costs and the other terms are variable costs. Because the costs increase with an increasing number of supplied products, we define $c_{i0}, c_{i1}, c_{i2} \geq 0$. (Note that the classic theory of the firm expands the costs usually up to the third order.)

Private economic activity occurs only when the business units (products) can realize a profit. In order to make profit, the nominal product price must increase the costs per unit given by:

$$c_i(s_i) = \frac{c_{i0}}{s_i} + c_{i1} + c_{i2}s_i \quad (7)$$

Note that the costs per unit have a minimum at an optimal output:

$$s_{i0} = \sqrt{\frac{c_{i0}}{c_{i2}}} \quad (8)$$

We want to denote this optimal output as the capacity limit. It corresponds approximately to the point of maximum productivity. We assume that business units try to work always close to the capacity limit in order to maximize their profit per unit.

*The demand side*

It is an empirical fact that the personal annual income distribution, $P(h)$ shows a two-class structure [12]. The upper class can be described by a Pareto power-law distribution. The majority of the population, however, belongs to the lower class. We want to separate the market potential into an upper and lower class contribution:

$$M = M_L + M_U \quad (9)$$

where the relative market potential in the USA is $m_U=M_U/M \approx 1-3\%$ and $m_L=M_L/M \approx 97\%-99\%$ of the population [13].

Further we introduce the market volume $V(\langle p \rangle)$, which is a function of the mean price. It determines the number of those agents who have sufficient personal annual income, $h$, to afford the good. The market volume is therefore the number of potential first purchase consumers, also denoted as potential adopters. In order to determine $V(\langle p \rangle)$ we make two general assumptions:

i) The upper class can always afford the good and is not limited by the product price.

Hence the market volume consists of the upper class and a part from the lower class $V_L(p)$, which depends on the mean price:

$$V(\langle p \rangle) = V_L(\langle p \rangle) + M_U$$
(10)

For the case that the good has also industrial applications we treat firms as agents not limited by the price, contributing to $M_U$.

As to evaluate $V(\langle p \rangle)$, we have to take advantage form the income distribution of the lower class. This distribution can be described by an exponential Boltzmann-Gibbs distribution, a lognormal distribution or a $\Gamma$-distribution (except for zero income) with an appropriate choice of the free parameters [14]. We confine here to the case that the income distribution of the lower class can be approximated by a Boltzmann-Gibbs distribution. In this case the relative abundance to find a representative agent having an annual income between $h$ and $h+dh$, can be given by the probability density function (pdf):

$$P(h) = \frac{1}{I} \exp(-h/I)$$
(11)

where the average personal income can obtained from:

$$I = \int_0^\infty dh P(h) h$$
(12)

ii) The key idea to evaluate the market volume is to assume that the chance to find an agent of the lower income class, who is willing to purchase the good, has a maximum at a minimum mean price.

Following Adam Smith we want to denote this price as the natural price $p_n$. It is determined on the one hand by the mean preferences and on the other hand it is a function of the personal income. In order to exclude the impact of the latter effect, we scale the mean nominal price $p_i$ by the mean income (of the lower class) and introduce the so called real price:

$$\mu_i = \frac{p_i}{I}$$
(13)

For a consistent notation, the model is essentially developed in terms of real prices. The second assumption also implies that for a mean price $\langle\mu\rangle \leq \mu_m$, the market volume must be equal to the market potential:

$$V(\mu_m) = M$$
(14)

The lower class contribution to the market volume can therefore be written as:

$$V(\langle\mu\rangle) = M_L \int_{\xi(\langle\mu\rangle)}^{\infty} P(z')dz'$$
(15)

where $z'=h/I$. While the integral determines the probability to find an agent with sufficient income, the unknown function $\xi(\langle\mu\rangle)$ determines how this probability varies as a function of the mean price.

The function $\xi(\langle\mu\rangle)$ can be specified by two conditions.
1. Because the cumulative income distribution is normalized to one, the function must be zero at $\mu_n$, in order to fulfil Eq.(14). Hence, we can approximate the function close to $\mu_n$ by a Taylor expansion. Up to the second order we can write for $\langle\mu\rangle > \mu_n$:

$$\xi(\langle\mu\rangle) \cong \xi_1(\langle\mu\rangle - \mu_n) + \xi_2(\langle\mu\rangle - \mu_n)^2$$
(16)

where $\xi_1, \xi_2 \geq 0$.

2. We demanded in *ii)* that the chance to find a potential consumer as a function of the price has a maximum at $\mu=\mu_n$. This condition implies that $\xi(\langle\mu\rangle)$ has a minimum at $\mu_n$ and therefore $\xi_1=0$.

The market volume can therefore approximated near $\mu_n$ by:

$$V(\langle\mu\rangle) \cong M_L \exp\left(-\frac{(\langle\mu\rangle - \mu_n)^2}{2\Theta^2}\right)$$
(17)

where $\Theta^2=1/(2\xi_2)$ is a constant for a market. The market volume scaled by the market potential gives the density:

$$v(\langle\mu\rangle) = \frac{V(\langle\mu\rangle)}{M} = v_L(\langle\mu\rangle) + m_U$$
(18)

## 2.2. The Market Dynamics

The market dynamics is determined on the one hand by the dynamics of supply and demand and on the other hand by the purchase process. It is an empirical fact that in everyday life the price varies slowly. In standard microeconomics this characteristic is known as price rigidity [15]. We want to take advantage from this property and introduce a separation of the time scales. On a short time scale the mean price is treated as slowly varying, such that the mean price can be considered to be nearly constant:

$$\frac{d\langle\mu\rangle}{d\tau} \sim \varepsilon \approx 0$$
(19)

with $\varepsilon<<1$. (When $t$ is in years and $\tau$ in weeks, $\varepsilon\approx 1/50$). The short time scale $\tau$ is related to the long time scale by:

$$t = \varepsilon\tau$$
(20)

Because the mean income $I$ can be treated as constant on the short time scale, nominal and real price variations are equivalent.

*The demand side dynamics*

The demand side is determined by potential consumers (agents), who want to purchase the good. The density of potential consumer's $\psi(t)$ is governed by the balance of their creation and disappearance. The creation rate of potential consumers is the demand rate $d(\langle\mu\rangle)$. It determines the number of potential consumers created per unit time scaled by the market potential. It consists of first purchase and repurchase demand. The first purchase demand is given by new potential adopters. In order to keep the model simple we confine here to consumer goods where the main number of agents correspond to the lower income class:

$$v_L(\langle\mu\rangle) = \frac{V_L(\langle\mu\rangle)}{M} >> m_U$$
(21)

(The reason for this assumption is that in this case Bass diffusion can be neglected. Bass diffusion is essentially due to the word-of-mouth effect which creates a sales wave for large $m_U$. This effect complicates the equations but is not necessary for an understanding of the firm evolution. For a complete study see [1]). In this case first purchase demand can be given by:

$$d_f(\langle\mu(t)\rangle) = \frac{dv(\langle\mu(t)\rangle)}{dt}$$
(22)

Repurchase demand must be proportional to the number of previous adopters of a good, given by the market volume $v(\langle\mu\rangle)$. The repurchase demand rate becomes:

$$d_r(\langle\mu(t)\rangle) = q(t)v(\langle\mu(t)\rangle)$$

(23)

where the rate *q(t)* is an average over the number of adopters. The total demand is determined by $d(\langle\mu\rangle)=d_f(\langle\mu\rangle)+d_r(\langle\mu\rangle)$.

On the short time scale is $d=d_r$, because the mean price and hence the market volume is a constant. Hence potential consumers appear with $d_r$ and disappear by purchasing the good with the total purchase rate $y_t$. Therefore the balance becomes:

$$\frac{d\psi}{d\tau} = d(\langle\mu\rangle) - y_t$$

(24)

The stationary density of potential consumers $\psi_S$ is determined by the condition $d\psi/d\tau=0$. Thus, in the stationary state the demand rate is related to the unit sales for a given mean price by:

$$y_t = d(\langle\mu\rangle)$$

(25)

*The supply side dynamics*

The key process on the supply side is the production and distribution of goods by firms (business units). However, the unit sales fluctuate on the short time scale. In order to compensate these fluctuations firms have inventories. We want denote the number of all available products of the *i-th* brand in inventories as $Z_i$, and consider stores also as inventories. (In this view stores are inventories of business units managed by dealers.) The total number of units scaled by the market potential is the density $z_i$. The density of available products can be obtained on the short time scale from the balance between supply and purchase flow:

$$\frac{dz_i}{d\tau} = s_i - y_i = \gamma_i y_i$$

(26)

where $\gamma$ is denoted as reproduction coefficient. A positive reproduction parameter expresses the degree of an excess supply, since $\gamma=0$ implies that demand equals supply. The total supply flow can be written as:

$$s_t = [1+\langle\gamma\rangle]y_t$$

(27)

while $\langle\gamma\rangle$ is the mean reproduction coefficient, i.e. the mean excess supply.

*The purchase process*

The key idea to model the purchase process is to consider the purchase of a good as a statistical event, where a potential consumer meets available units of the *i-th* brand and purchase them with a certain probability. The unit sales $y_i$ must be zero if there are either no potential consumers or available units. Expanding the unit sales of the *i-th* brand up to the first order, $y_i$ must be proportional to the product of both densities. Hence, purchase events occur with a frequency:

$$y_i \cong \eta_i z_i \psi(\mu_i)$$
(28)

where the rate $\eta_i > 0$ specifies the success of the *i-th* product and is denoted as preference parameter. This parameter is characterized by the product features (utility) and the (spatial) accessibility of a brand. The product price $\mu_i$ for which the brand is available limits the density of potential consumers $\psi(\mu_i)$. Expanding the price around the mean price:

$$\mu_i(\tau) = \langle \mu \rangle + \delta\mu_i(\tau)$$
(29)

the density of potential consumers can be expanded as

$$\psi(\mu_i(\tau)) = \psi(\langle \mu \rangle) + \delta\psi_i(\mu_i, \tau)$$
(30)

Taking the sum over all products, we obtain from Eq.(28):

$$y_t = \left(\psi(\langle \mu \rangle) + \delta\psi\right)\langle \eta \rangle_z z_t$$
(31)

where $z_t = \Sigma z_i$ and we approximated $\delta\psi_i \approx \delta\psi$. The brackets with index $z$ indicate the average over z. Applying this relation in Eq. (24) the time evolution of potential consumers becomes:

$$\frac{d\delta\psi}{d\tau} = -\delta\psi \langle \eta \rangle_z z_t$$
(32)

Since the mean preference parameter and $z_t$ are positive, fluctuations of the density of potential consumers always disappear. However, this occurs slowly when $z_t \to 0$, i.e. when the number of available products is very small, which is the case for catastrophes (e.g. war, earth quake etc.). We want to confine our considerations here to a functioning market economy and treat therefore fluctuations of the density of potential adopters always as small.

For the stationary state, we obtain:

$$y_t = \psi(\langle \mu \rangle)\langle \eta \rangle_z z_t = d(\langle \mu \rangle)$$
(33)

*The market equilibrium*

Because the mean price and hence the total demand rate $d(<\mu>)$ is a constant on the short time scale, Eq. (33) suggests that also the total sales must be a constant. Therefore:

$$\frac{dy_t}{d\tau} = \psi(\langle \mu \rangle)\langle \eta \rangle_z \langle \gamma \rangle y_t \sim \varepsilon \approx 0$$

(34)

where we used Eq.(26). Since $\psi(\langle\mu\rangle)$, $\langle\eta\rangle_z$, $y_t > 0$, Eq.(34) can be satisfied only when:

$$\langle\gamma\rangle \sim \varepsilon$$
(35)

Applying this result in Eq.(27) we obtain that total supply must be nearly equal to total demand at mean price:

$$s_t \cong d(\langle\mu\rangle)$$
(36)

This is the basic statement of the neo-classic theory. When this condition is fulfilled a market called to be in equilibrium [15].

We can establish the (short term) equilibrium demand curve from Eq.(23) to be

$$d(\langle\mu(t)\rangle) = q(t)\exp\left(-\frac{(\langle\mu(t)\rangle - \mu_n)^2}{2\Theta^2}\right)$$
(37)

which can be expanded near $\mu_n$ as:

$$d(\langle\mu(t)\rangle) = q(t) - \left(\frac{\langle\mu(t)\rangle - \mu_n}{\Theta}\right)^2$$
(38)

*The evolutionary dynamics*

When the total sales are constant in market equilibrium, the products (business units) are in competition for potential consumers. This leads to an evolutionary dynamics derived in this section.

Let us write the density of available products on the short time scale as:

$$z_i(\tau) = \langle z\rangle_N + \delta z_i(\tau)$$
(39)

where the mean value is averaged over the number of products indicated by brackets with index $N$:

$$\langle z\rangle_N = \frac{z_t}{N}$$
(40)

Than the time evolution of the unit sales of the *i-th* business unit can be obtained from a time derivative of Eq.(28):

$$\frac{dy_i}{d\tau} = \langle z \rangle_N \frac{d}{d\tau}(\psi(\mu_i)\eta_i) + \psi(\mu_i)\eta_i\gamma_i y_i$$
(41)

where we used Eq.(26). The first term represents the direct impact of price and preference fluctuations on the unit sales. However, it can be expected that consumer preferences very slowly, such that $d\eta/d\tau \approx 0$. As assumed above the density of potential consumers relaxes fast upon a price variation according to Eq.(32). Therefore $\delta\psi/d\tau \approx 0$, and we can neglected the impact of the first term in Eq.(41).

In market equilibrium the constraint Eq.(34) can be satisfied by adding a constant growth rate $\zeta$ such that:

$$\frac{dy_i}{d\tau} = (f_i - \zeta)y_i$$
(42)

while we have introduced the function:

$$f_i = \psi(\mu_i)\eta_i\gamma_i$$
(43)

Evaluating the sum over all products we obtain:

$$\zeta = \langle f \rangle = \frac{\sum_i y_i f_i}{y_t}$$
(44)

Rewriting Eq.(42), the sales evolution of the *i-th* model is determined by the replicator equation:

$$\frac{dy_i(\tau)}{d\tau} \cong (f_i - \langle f \rangle)y_i(\tau)$$
(45)

The parameter $f_i$ in the replicator equation is known as the fitness. Therefore we want to denote $f_i$ as the product fitness. This result implies that products suffer from an evolutionary competition. The sales of those brands with higher product fitness than the mean fitness increase and vice versa. The fitness space is essentially determined by the preference parameter $\eta$, the demand rate as a function of the product price $\mu$ and the reproduction coefficient $\gamma$. The fitness has therefore contributions from both sides of the market. Working conditions, environment pollution etc. are not contained in the fitness and have to be imposed externally. Note that it is just the condition of a market equilibrium that leads to the evolutionary replicator dynamics. A constant fitness advantage of a brand with respect to all other products changes the replicator equation into a Verhulst equation, which suggests a replacement of the other products according to a logistic law. For further discussion see [1].

The unit sales can be written in terms of the growth rate $r$ as:

$$r_i(\tau) = \frac{1}{y_i}\frac{dy_i}{d\tau} = \delta f_i(\tau)$$

(46)

with

$$f_i = \langle f \rangle + \delta f_i$$

(47)

and $\langle f \rangle \sim \varepsilon$. Hence, the replicator equation describes a multiplicative process with a growth rate given by the fluctuations of the product fitness, $\delta f$. Note that Eq.(46) expresses Gibrat's law of proportionate effects [4,11]. This law is a direct consequence of the competition between products in market equilibrium.

### 2.3. The Size Distribution

We have to distinguish between the size distribution of products and firms.

*The product size distribution*

The size distribution of the business units $P(y)$, is determined by the probability to find the unit sales of a business unit $y_i$ in the interval $y$ and $y+dy$. As shown above the unit sales of a product are determined by a stochastic multiplicative process. The size distribution can be given for the case that $\delta f$ can be treated as an independent, identical distributed, random (i.i.r.) variable. The central limit theorem suggests that in this case the size distribution of the business units is given for a sufficiently long time by a lognormal probability distribution function (pdf) of the form:

$$P(y,t) = \frac{1}{\sqrt{2\pi t}\,\omega y}\exp\left(-\frac{(\ln(y/y_0) - ut)^2}{2\omega^2 t}\right)$$

(48)

where $u$ and $\omega$ are free parameters and $y/y_0$ is the size of the business unit scaled by the size at $t=0$.

*The firm size distribution*

The size distribution of business firms $P(x)$, is given by the probability to find the unit sales of a firm $x_j$ in the interval $x$ and $x+dx$. As discussed above firms usually consist of several business units. Therefore the size distribution of firms will deviate from the lognormal distribution of their products. In order to derive the firm size distribution we want to establish a relation for the time evolution of firms.

On the one hand the firm sales are determined by the sum over the sales of their products. The growth of the firm sales is given therefore by the time derivative of Eq. (4):

$$\frac{dx_j(\tau)}{d\tau} \sim \sum_i r_i y_i(\tau)$$
(49)

where we used Eq.(46).

On the other hand, firms have also the ability to growth by adding new products. This can be done by creating new products, but also by mergers and acquisitions. Both processes are associated with considerable financial costs. Large firms, however, have a better chance to growth by these processes than smaller ones. In other words, the ability to growth is a direct function of the firm size. There are other factors that also support the growth of large firms as for example the market power and synergy effects between business units. Following previous research on firm growth, we want to denote this size dependent growth as preferential attachment. The key idea to take preferential attachment into account is to add a small size dependent contribution $F(x)$ to Eq.(49):

$$\frac{dx_j(\tau)}{d\tau} = F(x_j) + \sum_i r_i y_i(\tau)$$
(50)

If the firm size is zero, there is no preferential attachment. Therefore the size dependence can be expanded up to the first order as:

$$F(x) \cong Ax$$
(51)

where the rate $A$ is a positive free parameter.

In order to solve Eq.(50) we want to make two approximations:

1. It is known from empirical studies that most firms have a superior product, which is called "cash cow" [16]. It is the main source of income of a firm. We want to apply the cash cow concept and approximate the sum over all products by its main contribution:

$$\sum_i r_i y_i \cong r' y' \cong r' G(x_j)$$
(52)

where $y'$ are the sales of the cash cow and $G(x_j)=x_j$.

2. Further the fitness fluctuations around the mean fitness are treated as white noise on the short time scale, such that the growth rate of the cash cow $r'=\rho$ is an i.i.d. random variable, with mean value and time correlation function:

$$\langle \rho(\tau) \rangle_\tau = 0$$

$$\langle \rho(\tau), \rho(\tau') \rangle_\tau = 2D\delta(\Delta\tau)$$

(53)

The brackets with index $\tau$ indicate the time average, $D$ is a noise amplitude and $\Delta\tau=\tau-\tau'$. With these approximations the evolution of the firm sales turns into a generalized Langevin equation:

$$\frac{dx}{d\tau} = F(x) + \rho G(x)$$

(54)

It can be solved to give for a sufficiently long time a size distribution of the form (Appendix A):

$$P(x) \sim \frac{1}{x^{\left(1+\frac{A}{D}\right)}}$$

(55)

This result suggests that the firm size distribution in terms of unit sales approaches a power law (Pareto distribution) which can be related to Zipf's law [17].

However, mergers, acquisitions, the addition of new products etc. are rare events on the short time scale. Therefore the preferential attachment mechanism is assumed to be small ($A\sim\varepsilon$). For small firms ($x\sim\varepsilon$) this mechanism can therefore be neglected. The firm size distribution for small firms is therefore approximately given by a lognormal size distribution of the unit sales of the main product, $P(x)\approx P(y')$. For large firms ($x>>\varepsilon$), the preferential attachment mechanism transforms the firm size distribution into a power law. Hence, the size distribution of firms is suggested to be a lognormal distribution with a power law tail.

## 2.3. The Price Distribution

As mentioned above, business units are constrained on the short time scale by a limited capacity. Therefore they have to respond on large sales variations by varying the product price. In order to be close to the capacity limit we assume that business units have the tendency to increase the price when the unit sales considerably increase in time and vice versa. This rule can be formulated as follows:

$$\frac{d\delta\mu}{d\tau} \sim sign\left(\frac{dy}{d\tau}\right)$$

(56)

where we used that the mean price is a constant on this time scale. Scaling within the sign-function by the positive variable $y$, we can write:

$$\frac{d\delta\mu}{d\tau} \sim sign\left(\frac{1}{y}\frac{dy}{d\tau}\right) \sim sign(\delta f(\delta\mu))$$

(57)

while we have applied the replicator equation.

Hence, the model suggests that short term price variations are the result of fitness fluctuations of the products around the mean fitness. Expanding the product fitness around the mean price:

$$\delta f(\delta\mu) = \frac{\partial f(\langle\mu\rangle)}{\partial\mu}\delta\mu$$
(58)

we obtain for the evolution of the price fluctuations:

$$\frac{d\delta\mu}{d\tau} \sim sign\left(\frac{\partial f(\langle\mu\rangle)}{\partial\mu}\delta\mu\right)$$
(59)

With $f(\langle\mu\rangle)=\langle f\rangle$ and Eq. (43) we can write:

$$\frac{\partial f(\langle\mu\rangle)}{\partial\mu} = \langle\gamma\rangle\langle\eta\rangle\frac{\partial\psi(\langle\mu\rangle)}{\partial\mu} \sim \varepsilon\frac{\partial d(\langle\mu\rangle)}{\partial\mu} < 0$$
(60)

where we used Eq.(33).

Since the fitness derivation is negative, the tendency of the business units to minimize the costs as given by Eq.(56) can be interpreted as a restoring force:

$$\Phi(\delta\mu) \sim -sign(\delta\mu)$$
(61)

driving the price of the individual products back towards the mean price. Writing the restoring force as due to a generalized potential V':

$$\Phi(\delta\mu) = -\frac{dV'(\delta\mu)}{d\delta\mu}$$
(62)

and treating deviations from the mean as small stochastic fluctuations $\zeta(\tau)$, Eq.(57) turns into a Langevin equation of the form:

$$\frac{d\delta\mu}{d\tau} = -\frac{dV'}{d\delta\mu} + \zeta$$
(63)

As a first approximation we want to consider the short term fluctuations as uncorrelated white noise with mean value und time correlation:

$$\langle\zeta(\tau)\rangle_\tau = 0$$

$$\langle \zeta(\tau), \zeta(\tau') \rangle_\tau = D'\delta(\Delta\tau)$$

(64)

The stationary price distribution $P(\delta\mu)$ defined by the probability to find a price variation of a sold product $\delta\mu$ in the interval $\mu$ and $\mu+d\mu$, turns into:

$$P(\delta\mu) \sim \exp\left(-\frac{|\delta\mu|}{\sigma_\mu}\right)$$

(65)

(For an explicit derivation by applying the corresponding Fokker-Planck equation see [1]). The distribution of price fluctuations in a competitive market is therefore given for uncorrelated fluctuations by a Laplace (double exponential) distribution, where $\sigma_\mu$ is the standard deviation. In a semi-log plot, the Laplace distribution has a tent shape around the mean price.

The point is, however, that the purchase process is a human activity. But human activities are not uncorrelated. Empirical investigations suggest that human activities occur in bursts. The time correlation for human activities is governed by a long term correlation function of the form [18]:

$$\langle \zeta(\tau), \zeta(\tau') \rangle_\tau \sim \Delta\tau^{-\nu}$$

(66)

decaying as a power law with an exponent $0<\nu<1$. Therefore, price fluctuations are not uncorrelated but have instead a pronounced mountain-valley structure, where statistically large values are likely to be followed by large values and small by small ones.

The main difference to uncorrelated fluctuations is that the standard deviation of the price becomes a function of the size of the business unit's $\sigma_\mu = \sigma_\mu(y)$, with [18-20]:

$$\sigma_\mu(y) \sim y^{-\beta}$$

(67)

and $0<\beta<1/2$. Applying this result to Eq. (65), we have to introduce a conditional price distribution of the products:

$$P(\delta\mu|y) \sim \exp\left(-\frac{|\delta\mu|}{\sigma_\mu(y)}\right)$$

(68)

The exponent of the standard deviation $\beta$ and the correlation function $\nu$ are linked by the scaling relation [18]:

$$\beta = \frac{v}{2}$$

(69)

The size independent total price distribution turns into:

$$P(\delta\mu) = \int_0^\infty P(\delta\mu|y) P(y) dy$$

(70)

where the size distribution $P(y)$ is given by the lognormal distribution Eq.(48) and the conditional price distribution by Eq. (65). In order to obtain an approximate solution for the total price distribution, we want to reduce the lognormal size distribution to its main contribution. For sufficiently large sales $y \geq y_m$, the size distribution of the products can be represented by its tail, which decays as $P_y \sim y^{-1}$. This approximation yields:

$$P(\delta\mu) \sim \int_{y_m}^\infty \frac{1}{\sigma_\mu(y)} \exp\left(-\frac{|\delta\mu|}{\sigma_\mu(y)}\right) \frac{dy}{y}$$

(71)

while $\sigma_\mu(y)$ is given by Eq. (67). The integration starts at a minimum size $y_m$, at which the size distribution can be given by its tail. Carrying out the integration we obtain that the price distribution can be approximated for $|\delta\mu|>0$ by:

$$P(\mu) \cong C_\mu \frac{\exp\left(-\frac{|\mu - \langle\mu\rangle|}{\sigma_{\mu m}}\right)}{|\mu - \langle\mu\rangle|}$$

(72)

where $\sigma_{\mu m} = \sigma_\mu(y_m)$, $C_\mu$ is a normalization constant and we used Eq. (29).

## 2.4. The Mean Price Evolution

Writing Eq.(2) in a continuous form the evolution of the mean price is determined by:

$$\frac{d\langle\mu\rangle}{d\tau} = \int_0^\infty \frac{dP(\mu')}{d\tau} \mu' d\mu' = \int_0^\infty P(\mu') \mu' (f(\mu') - \langle f \rangle) d\mu'$$

(73)

where we used the replicator equation. This relation turns into:

$$\frac{d\langle\mu\rangle}{d\tau} = \int_0^\infty \mu' f(\mu')P(\mu')d\mu' - \langle\mu\rangle\langle f\rangle$$
(74)

and with Eq.(58) we obtain on the long time scale:

$$\frac{d\langle\mu\rangle}{dt} = \frac{1}{\varepsilon}\frac{df(\langle\mu\rangle)}{d\mu}Var(P(\mu))$$
(75)

where the price variance is defined as:

$$Var(P(\mu)) = \int P(\mu')\mu'^2 d\mu' - \left(\int P(\mu')\mu' d\mu'\right)^2$$
(76)

Using Eq.(60) and Eq.(38) we get:

$$\frac{1}{\varepsilon}\frac{\partial f(\langle\mu\rangle)}{\partial\mu} \sim \frac{\partial d(\langle\mu\rangle)}{\partial\mu} \sim \langle\mu(t)\rangle - \mu_n$$
(77)

Integrating Eq.(75) the mean price decreases in time approaching $\mu_n$ according to:

$$\langle\mu(t)\rangle = \mu_0 e^{-at} + \mu_n$$

(78)

where $\mu_0$ is the price at $t=0$ and the parameter $a$ is denoted as the price decline rate [1]. This result applies only, when $a\sim Var(P(\mu))>0$, i.e. a monopoly market with $a=0$ is excluded from the model. The long term price decline is a direct consequence of the competition between the brands.

Note that the mean reproduction parameter as governed by Eq.(27) can be rewritten in equilibrium as:

$$\langle\gamma\rangle = \frac{s_t(\langle\mu\rangle)}{d(\langle\mu\rangle)} - 1$$
(79)

While on the short time scale $<\gamma>$ is of the order $\varepsilon$, on the long time scale this is not the case. The mean price evolution can be determined by rewriting Eq.(75), taking advantage from Eq.(60) and Eq.(79) to get:

$$\frac{d\langle\mu\rangle}{dt} = \frac{\langle\eta\rangle Var(P_\mu)}{d(\langle\mu\rangle)\langle\eta\rangle_z z_t}\left|\frac{\partial d(\langle\mu\rangle)}{\partial\mu}\right|(d(\langle\mu\rangle) - s_t(\langle\mu\rangle))$$
(80)

The neo-classic Walrasian theory explains the mean price evolution as a tatonnement process. It suggests that the mean price is determined by the relation [21]:

$$\frac{d\langle\mu\rangle}{dt} = H(d(\langle\mu\rangle) - s_t(\langle\mu\rangle))$$
(81)

with $H>0$.

Comparing both results the mean price evolution of the evolutionary model is formally equivalent to the Walrasian picture. Both suggest that the mean price increases when total demand increases total supply and vice versa. The Walrasian approach suggests, however, that the mean price is a stable state only when aggregate demand is equal to aggregate supply. In the evolutionary picture, though, the distribution of price fluctuations is governed by Eq.(72). It is always stable as long as $\partial f/\partial\mu<0$, reverting the mean price to the natural price according to Eq.(65). However, when demand increases supply caused by an internal or external event, the price distribution becomes unstable, since $\partial f/\partial\mu>0$, accompanied with an increase of the mean price in agreement with the Walrasian picture.

The evolutionary model suggests therefore that all states with competition, where total demand (slightly) increases total supply, are equilibrium states. A mean (real) price increase is caused by the absence of competition. Note that this is even the case when demand is equal supply, i.e. when the neo-classic model assumes equilibrium. Whether such a situation actually occurs depends on the characteristics of the good. The purchase of durable goods for example can be shifted in time in the case of a supply shortage. Therefore for durable goods a mean price increase caused by a supply shortage is rather unlikely (except for catastrophic events). On the other hand for non-durables (e.g. food, oil), mean price jumps are very likely [22]. Note that a nominal price increase on the long time scale due to an increased income (inflation) is not considered here.

## 2.5. The Growth Rate Distribution

We want to consider the growth rate distribution of products. The growth rate distribution $P(r)$, determines the relative abundance of a business unit to have a growth rate in the interval $r$ and $r+dr$. The growth rate dynamics is determined on the short time scale by Eq.(46) and can be approximately written as:

$$r \cong \log\left(\frac{y(\tau+\Delta\tau)}{y(\tau)}\right)$$
(82)

The replicator dynamics suggests that growth rate fluctuations are caused by fitness fluctuations. We want to consider short term growth rate fluctuations as essentially due to price fluctuations because the price can be varied easily compared to the other parmeters determining the fitness. We write:

$$r \cong \delta f(\mu) = \frac{\partial f(\langle\mu\rangle)}{\partial\mu}\delta\mu$$
(83)

In this approximation price fluctuations are directly related to growth rate fluctuations. Because the growth rate fluctuations are correlated, the growth rate distribution must be a function of the firm size and can be obtained from Eq.(68) by changing variables according to Eq. (83):

$$P(r|y) \sim \exp\left(-\frac{|r|}{\sigma_r(y)}\right)$$
(84)

Here we used that the mean growth rate on the short time scale is:

$$\frac{1}{y_t}\frac{dy_t}{d\tau} = \frac{1}{y_t}\sum_i r_i y_i = \langle r \rangle \sim \varepsilon$$
(85)

The size dependence of the standard deviation is equivalent to price fluctuations governed by:

$$\sigma_r(y) \sim y^{-\beta}$$
(86)

Taking advantage from the approximation leading to Eq.(72) we can conclude that the total growth rate distribution for $|r|>0$ has the form:

$$P(r) \cong C_r \frac{\exp\left(-\frac{|r|}{\sigma_{rm}}\right)}{|r|}$$
(87)

where $\sigma_{rm}=\sigma_r(y_m)$ and $C_r$ is a normalization constant.

**2.6. The Long Term Market Evolution**

Finally we are able to establish some relationships concerning the long term evolution of the market.

*The product life cycle*

The product life cycle determines the total sales as a function of the time. Because the market evolves by running through quasi equilibrium states, the total sales are given by Eq.(25):

$$y_t(t) = y_f(t) + y_r(t) = \frac{dv(\langle\mu(t)\rangle)}{dt} + q(t)v(\langle\mu(t)\rangle)$$
(88)

The evolution of the first purchase sales $y_f(t)$ can be obtained by inserting Eq.(78) in Eq.(17) for the market volume and perform the time derivative. We get:

$$y_f(t) = 2a\kappa n(t)\exp(-2at)$$
(89)

where

$$\kappa = \left(\frac{\mu_0}{2\Theta^2}\right)^2$$
(90)

and

$$v(t) = n(t) = n_0 \exp\left(-ke^{-2at}\right)$$

(91)

while $n(t)$ can be interpreted as the adopter density with $n_0=1$. The evolutionary model suggests therefore that there exists a diffusion process caused by the expansion of the market volume with a decreasing price. It is determined by Gompertz equation (Eq.(91)) and therefore denoted as Gompertz diffusion.

The repurchase sales $y_r(t)$ is the sum of replacement purchase $y_R(t)$ and multiple purchase $y_m(t)$ such that:

$$y_r(t) = y_m(t) + y_R(t)$$

(92)

Since, multiple purchase must be proportional to the market volume, $v(t)$, the unit sales can be approximated by:

$$y_t(t) = qv(t)$$
(93)

where $q>0$ is a mean multiple purchase rate.

Replacement purchase has to be taken explicitly into account only when the mean life time of a good $t_p \gg 1/a$. In this case the correlated first purchase process (diffusion) leads to a correlated repurchase wave. This wave occurs with a periodicity given by the mean lifetime of the good. In the simplest approximation replacement purchase induces periodic variations of the first purchase wave for $t>t_p$ given by:

$$y_R(t) \cong \chi y_f(t - t_p)$$
(94)

else, $y_R(t)=0$, while $\chi>0$ is the fraction of previous sales suffered from replacement purchase. In the standard theory of economic fluctuations these waves are known as Juglar cycles with a periodicity of about 8-10 years [1, 23]. Since non-durables have a short lifetime, Juglar cycles are relevant only for durables. Note that the short time scale is characterized by multiple purchase as was used in Eq.(23), while Juglar waves leads to a variation of $q(t)$.

*The learning curve*

We have to emphasize that a business unit cannot sell products below the production costs. Therefore the product price must increase for increasing manufacturing costs per unit. (This leads to the so called supply curve in the classic model). Thus the costs per unit can be viewed to be a function of the product price. We take advantage from this relationship and expand the costs per unit up to the first order as a function of the price:

$$c_i(\mu_i) = \alpha_i(t)\mu_i(t)$$

(95)

Since the profit per unit is always positive the coefficient is in the range $0<\alpha_i(t)\leq 1$. We write the coefficient as the sum of the average over all products $<\alpha>_N$ and time dependent fluctuations:

$$\alpha_i(t) = \langle\alpha\rangle_N + \delta\alpha_i(t)$$

(96)

Because $<\alpha>_N <1$, fluctuations must be even smaller $\delta\alpha(t)<<1$. Therefore we can neglect the time dependent fluctuations compared to the mean value. Taking the average over Eq.(95) the evolution of the mean costs per unit can be written as:

$$\langle c(t)\rangle \cong \langle\alpha\rangle_N \langle\mu(t)\rangle$$

(97)

Since the mean price evolution is given by Eq.(78), the costs per unit decrease in time. This effect is known in economic literature as the learning (experience) curve [24,25]. And indeed the adaptation process induced by the mutual competition between the products can be considered as a learning process.

However, following Ebbinghaus the learning processes have usually the form of a power law:

$$\langle c(t)\rangle \sim w(t)^{-\beta}$$

(98)

where $w(t)$ is the cumulative output of the business units and $\beta$ is the elasticity of costs with regard to output. This relationship is denoted in economic literature as Henderson's law [24]. Empirical investigations suggest a decline of the costs per unit of the order of 10-25% for a doubling of the cumulative output.

Therefore we have two statements for the mean costs. On the other hand the mean costs should have the form of Henderson's law Eq.(98) and on the other hand the evolutionary model suggests an exponential decrease of the costs according to Eq.(97).

*The law of diminishing returns*

From Eq.(5) using Eq.(97) we get the relation:

$$\langle g(t) \rangle = (1 - \langle \alpha \rangle_N) \langle \mu(t) \rangle$$
(99)

Since the mean price is governed by Eq. (78), also the profit margin (return per unit) exhibits an exponential decrease with time. This tendency was already anticipated by Adam Smith and is known in economic literature as the law of diminishing returns [26]. Eq.(99) can be rewritten as:

$$\frac{G_t(t)}{R_t(t)} = \frac{\langle g(t) \rangle y_t(t)}{\langle \mu(t) \rangle y_t(t)} = (1 - \langle \alpha \rangle_N)$$
(100)

In view of the fact that the right hand side of this relation is time independent, also the left hand side must be constant. This is a remarkable result, because it suggests that the total profit $G_t(t)$ scaled by the total revenue $R_t(t)$ (respectively total costs) is a time independent invariant of a market.

*The market size*

We want to characterize the market size by the number of active firms $N_f(t)$. The market size is confined by the condition that the total costs cannot be larger than the total revenue of a market. Writing the total costs as:

$$C_t(t) = \langle c(t) \rangle y_t(t)$$
(101)

we obtain with Eq.(97):

$$\frac{C_t(t)}{R_t(t)} = \langle \alpha \rangle_N \leq 1$$
(102)

Introducing the mean costs per firm as:

$$\langle\langle C(t) \rangle\rangle = \frac{1}{N_f(t)} \sum_{j=1}^{N_f} C_j(t) = \frac{C_t(t)}{N_f(t)}$$
(103)

where the double brackets indicate the average over the number of firms, the condition Eq.(102) leads to the number of firms:

$$N_f(t) = \langle \alpha \rangle_N \frac{R_t(t)}{\langle\langle C(t) \rangle\rangle}$$
(104)

The time evolution of the number of firms on the long time scale is therefore given by:

$$\frac{dN_f(t)}{dt} = \langle\alpha\rangle_N \left( R_t(t)\frac{d\langle\langle C(t)\rangle\rangle^{-1}}{dt} + \langle\langle C(t)\rangle\rangle^{-1}\frac{dR_t(t)}{dt} \right)$$
(105)

The key idea to determine the market size is that the mean production costs cannot adapt arbitrarily fast to demand variations (due to fixed costs). In periods of the product life cycle where the revenue varies much faster than the mean costs:

$$\frac{dR_t}{dt} \gg \frac{d\langle\langle C(t)\rangle\rangle^{-1}}{dt}$$
(106)

the evolution of the number of firms can be given approximately by the integration of Eq.(105), while the condition Eq.(106) can be satisfied by setting $d\langle\langle C(t)\rangle\rangle^{-1}/dt \approx 0$. Therefore, the number of business units is governed by the total revenue of the market:

$$N_f(t) \approx \frac{\langle\alpha\rangle_N}{\langle\langle C\rangle\rangle} R_t(t) + N_{f0} = B\langle\mu(t)\rangle y_t(t) + N_{f0}$$
(107)

while $B$ is a proportionality and $N_{f0}$ is an integration constant.

When the total revenue varies slowly, which is the case when the mean price is close to the natural price, we have the condition:

$$\frac{dR_t}{dt} \ll \frac{d\langle\langle C(t)\rangle\rangle^{-1}}{dt}$$
(108)

In this case the number of firms is approximately determined by the evolution of the mean costs:

$$N_f(t) \approx \frac{\langle\alpha\rangle_N}{\langle\langle C(t)\rangle\rangle} R_t + N'_{f0} \to N'_{f0}$$
(109)

Since the mean costs increase in time, the number of firms becomes a constant after sufficient time. Note that the period of a decreasing market size is known in economic literature as shakeout [27].

## 3. Comparison with Empirical Results

The evolutionary theory makes a number of predictions that can be tested or are already known. The theory suggests that:
1. The size (in terms of unit sales) of products is governed by a lognormal distribution.
2. Firms have a lognormal size distribution with a departure in the upper tail which decays as a power law.
3. Price fluctuations of the products are a function of the product size. The distribution of price fluctuations is approximately determined for a given size by a Laplace distribution. Its standard deviation is related to the size by an exponential law. The total price distribution is given by Eq.(72).
4. The growth rate distribution is like the price distribution a function of the size of the product (firm) and determined by a Laplace distribution for a given size. It exhibits the same size-variance relationship as the price distribution. The total growth rate distribution can be approximated by Eq.(87).
5. The mean price decreases according to an exponential law as long as there is competition between the products, i.e. when total supply increases total demand. The price distribution becomes unstable, when total demand increases total supply associated with an increase of the mean price.
6. The decrease of the mean price is associated with an increase of the market volume. This process can be interpreted as a diffusion process, which is governed by Gompertz equation and therefore denoted as Gompertz diffusion.
7. Because the product price reflects the costs per unit, they can be considered to be a function of the price (supply curve). This has two consequences. Since the mean price decreases in time, the mean costs must exhibit the same time dependence. This effect is known as the learning (experience) curve. Henderson's law suggests that the costs per unit decrease with the accumulated output according to a power law. The second effect is that when the mean price and the mean costs exhibit an exponential decrease, also the profit margin (profit per unit) must be governed by an exponential decrease. This is the so-called law of diminishing returns. It implies that the total profit scaled by the total revenue is an invariant even on the long time scale.
8. The number of firms in a market is confined by the condition that the total costs cannot be larger than the total revenue.

That the size distribution of large firms in terms of unit sales, revenue, employees, assets etc. can be approximated in the upper tail by a Pareto distribution is well known [11,17,28]. Empirical studies of the worldwide pharmaceutical industry also revealed the relationship between the size of products and firms [29,30]. They found that the size distribution of products is lognormal, while for firms the upper tail has the form of a power law distribution. This is in agreement with the presented model.

Further the model suggests that the price and growth rate distributions are a function of the size and can be approximated by a Laplace distribution for products. The cash-cow concept suggests that this is also the case for the growth of firms since they are determined by the growth of their main product. Caused by correlated fluctuations of human activity, the standard deviation of the size dependent Laplace distribution can be expected to decay as a function of the size as a power law with an exponent $\beta$. Empirical investigations of human activity suggests an exponent $\beta \approx 0.2$ [18-20]. That the growth rate of firms can be approximated by a Laplace distribution is well known [31-35]. Both growth rate distributions and exponents of the standard deviation are very similar, as suggested by the model. Empirical investigations deliver a $\beta \approx 0.17$ for firms and $\beta \approx 0.15$ for products [10,30].

The total growth rate and price distribution is often approximated in empirical investigations by a Subbotin distribution. For the growth rate it has the form [35]:

$$P(r) = C_1 \exp\left(-C_2 |r - \langle r \rangle|^\beta\right)$$
(110)

with constant coefficients. It is a normal distribution for $\beta=2$ and a Laplace distribution for $\beta=1$. The evolutionary model suggests that the total growth rate distribution can be approximately given by Eq.(87).

In order to show the applicability of the model we take as an example the empirical data for the total growth rate distribution obtained by De Fabritiis et al. [30] and fitted the data with a Subbotin distribution (dotted line) as given by De Fabritiis and by Eq. (87) (fat line) displayed in Figure 1. Both are almost equivalent, which indicates that Eq.(87) is a satisfactory approximation of the total growth rate distribution for $|r|>0$.

That the price distribution can be fitted by a Subbotin distribution was shown for example for spot prices in the electricity market [36,37]. That correlated price fluctuations induce a self similar mountain-valley structure in time was first established by Mandelbrot for commodities [38]. However, the variation of the price has in this model also an impact on the sales, respectively on the number of available products in inventories. These quasi-cyclic commodity price and inventory variations in the order of several months are known in economic theory as Kitchin-cycles [23,39]. The present model suggests that Kitchin-cycles are correlated fluctuations rather than cycles. The evolutionary model also predicts that price fluctuations must exhibit a size-variance relationship similar to the growth rate (Eq.(67)). An investigation of this relationship is not known to the author.

As an example of the long term evolution of a market, we compare the predictions of the model with empirical data available for the evolution of the US market of Black & White (B&W) TV sets as investigated by Wang [27]. Displayed in Fig.2 are the nominal price (triangle) and the market penetration (circles) of B&W TV's. The fat lines are applications of the present model with parameters given in [1]. As can be seen the mean price decreases according to an exponential law, while the market penetration can be described by Gompertz diffusion. Displayed in Fig.3 are the corresponding unit sales, which are the result of first and repurchase processes. First purchase is a combination of Gompertz and Bass diffusion not further specified here (see [1]). The model reflects qualitatively the periodic variations of the total sales as expected from the model (Juglar waves).

Displayed in Fig. 4 is the empirical mean price given in Fig.1 as a function of the empirical cumulative output. Also shown is a learning curve of the form Eq.(98) with $\beta=-0.32$, which expresses a 20% reduction of the mean price for every doubling of the cumulative output. As can be seen the empirical data of the mean price can be approximately described by Henderson's law. But Fig.2 suggests that the mean price is also given by an exponential decline. This result implies that the mean price dynamics is governed by the competition between the brands, reflecting the decreasing production costs due to a learning process.

Finally displayed in Fig.5 is the empirical number of firms in this durable market as given by Wang [27]. Eq.(107) suggests that the market size is essentially determined by the revenue in the beginning of the product life cycle, because the revenue varies considerably in time. The fat line in this figure is proportional to the empirical revenue obtained from the empirical unit sales in Fig.3 multiplied with the mean price from Fig.2. Up to the middle of the nineteen sixties the number of firms is roughly proportional to the market revenue. After this period the number of firms approaches a nearly constant value as expected by the model. This trend is according to this theory a result of the limited financial capacity of the TV-market.

## 4. Conclusion

Private economic activity occurs only, when a financial gain can be achieved. This goal implies a high productivity of the internal processes of a firm, associated with minimum costs. In order to keep close to the capacity limit, demand fluctuations are compensated by product price fluctuations towards the mean price. As a result the mean price of a market varies slowly. This compensation process leads to a market where total demand is nearly equal to total supply, which is denoted in economic literature as market equilibrium.

However, it is a quasi-equilibrium state (in the order of month). The model suggests that the compensation process works only when there is competition between products (business units), which implies a total excess supply. Although the neo-classic theory and the evolutionary model deliver a formally equivalent relation for the mean price evolution, they have different interpretations. While the neo-classic Walrasian picture suggests a single stable state when total demand is equal to total supply, in the evolutionary model the mean price is always stable when total supply increases total demand. In other words, in the Walrasian model a stable mean price is associated to just one state. In the evolutionary theory all states with competition are stable with a nearly constant mean price on the short time scale.

Because total demand is governed by the mean price, also the total sales are nearly constant within short time periods. This effect creates an evolutionary competition between the products in a free market. The mutual competition is therefore a direct consequence of the market equilibrium. The evolution of the individual sales is governed by a replicator dynamics, while the sales process is a function of the product fitness. The product fitness is essentially defined by an excess supply of the good, the preference (product utility) and the price. Gibrat's law is a result of this competition and can be understood as sales variations induced by fitness fluctuations.

Note that the driving force of evolutionary processes is a shortage. In the case of a product market it is the shortage of a sufficient number of potential consumers. Business units have an advantage, when they find more consumers than their competitors. This can be done by increasing the product fitness, i.e. a lower price, a higher preference and a higher presence in the market realized by an excess supply. The search of the business units for a competitive advantage can be viewed as a learning process while each firm has its own history. The competition between firms may lead to an "arms race" for the price (learning curve [39]) and in the variation of products, similar to species in predator-prey systems. The model implies that concerning the preference, the evolution of products can be visualizes as an evolutionary tree (but in difference to species they may have several roots). Concerning the mean price the market evolution can be predicted as long as there is competition. The market price decreases according to an exponential law approaching the natural price asymptotically. Associated with this price decrease is an extension of the market volume, which can be interpreted as a diffusion process. Since this diffusion is related to Gompertz equation, it is denoted as Gompertz diffusion.

Based on this approach, the product and firm size distributions are derived. While the size distribution of products is exclusively determined by Gibrat's law (fitness fluctuations), the firm size distribution is governed by two evolutionary processes: the preferential growth of the products governed by Gibrat's law and a preferential attachment of new products by successful firms. Both processes are recognized by previous investigations to dominate the firm growth. Note, however, that the presented model derives these processes directly from a consideration of the dynamics of a free market. The evolutionary model suggests that products (brands) have a lognormal size distribution, while the firm size distribution is lognormal for small firms but has a power law tail for large firms.

Since individual products are governed by a replicator dynamics, the growth rate of products and firms can be also derived. We have simplified the fitness fluctuations on the short time scale to be essentially determined by product price fluctuations, because the price can be varied easily. In view of the fact that fitness fluctuations are directly related to growth rate fluctuations the price distribution is the key to understand the growth rate distribution.

The price distribution is derived on basis that there is a restoring force for price fluctuations on the short time scale. The restoring force contains the tendency of the business units to work close to the point of minimum costs (capacity limit). For uncorrelated price fluctuations we obtain a Laplace distribution around the mean price. However, the purchase process can be viewed as a human activity. Human activity is known to be highly correlated. Taking this characteristic into account, the standard deviation of the Laplace distribution becomes a function of the size (unit sales). The standard deviation of the firms and products of the Laplace distribution is governed by a power law as a function of the size with exponent *β*. Form empirical investigations of human activity it is known that *β≈0.2*. The total price distribution turns into a Subbotin-like distribution of the form Eq.(72).

Because growth rate fluctuations are on the short time scale directly related to price fluctuations, the growth rate distribution is formally equivalent to the price distribution. Therefore the total growth rate distribution is also Subbotin-like given by Eq.(87), and the size-variance relationship of the Laplace distribution has the same exponent as human activity.

The comparison with empirical investigations suggests that the theory is able to explain the firm (product) size and growth. The picture drawn from the evolutionary model is in agreement with empirical data for the size and growth rate distribution of products and firms including the empirical size-variance relationship. Also the self similar structure of the time evolution of commodity and stock prices and their quasi periodic variations (Kitchin cycles) is well known. The model predicts, however, that even the price must be a function of the size with the same size-variance relationship that holds for growth rates.

The theory also derives the number of firms that can survive in a market, called market size. The model does not allow the determination of the bankruptcy risk of an individual firm [17,40]. Though, the market size can be approximately determined by the constraint that the total costs cannot increase total revenue over a long time period. In biological evolution species become extinct when the environment changes much faster than the evolutionary adaptation process can respond. In the same spirit the shakeout of firms is derived as due to changes in the total revenue of a market. A comparison with empirical results exhibits a good qualitative coincidence. Note that the model also suggests a remarkable invariance. The relation between total profits to total revenue, respectively to total costs, is a constant even on the long time scale. When the total profit is interpreted as capital income and the total costs as labour income, this invariance can be related to the empirical findings of Cobb and Douglas suggesting that the quotient of total capital to total labour income is a time-independent constant [41].

Note that a monopoly market or a market with fixed prices undergoes a limited evolution. Therefore a monopoly market or a command economy evolves much slower than free markets. We have to emphasize that the presented theory is a combination of an evolutionary and a standard neo-classic view of a market. While the neo-classic theory considers firms in market equilibrium, evolutionary models focus on entrepreneurship, the competition between firms within and between industrial sectors and how competitive advantages can be achieved [42,43]. According to the presented model is the key difference the time scale. While market equilibrium occurs within short time periods, evolutionary processes come into play on the long time scale, when the mutual competition between products (firms) is driving the fitness towards its maximum.

As found earlier, evolution has similarities with continental drift [44]. In every day life the earth crust is in a quasi equilibrium state. Only on a long time scale the drift of the continents is evident. We notice the motion of the continents only by large events.

Equivalently a market can be treated on a short time scale as in quasi equilibrium (punctuated equilibrium) disturbed by some extraordinary market events. On a long time scale we can see the evolution at work.

**Appendix A**

The firm sales Eq. (54) have the form of a generalized Langevin equation:

$$\frac{dx}{d\tau} = F(x) + \rho G(x)$$
(A1)

This multiplicative stochastic relation can be transformed into a relation with additive noise by introducing the functions [45]:

$$\frac{dh(x)}{d\tau} = \frac{1}{G(x)} \frac{dx}{d\tau}$$
(A2)

and

$$-\frac{dV(x)}{dh(x)} = \frac{F(x)}{G(x)}$$
(A3)

Inserting these relations into Eq. (A1) we obtain the Langevin equation:

$$\frac{dh}{d\tau} = -\frac{dV}{dh} + \rho$$
(A4)

For uncorrelated fluctuations this equation describes a random walk in the potential $V(x)$. For a sufficiently long time the probability distribution approaches:

$$B(h)dh = \frac{1}{N'} \exp\left(-\frac{V(H)}{D}\right) dH$$
(A5)

where $N'$ is a normalization constant. In terms of the original variable, we get:

$$P(x)dx = B(h)dh = \frac{1}{N'} \exp\left(-\frac{1}{D} \int \frac{F(x')}{G(x')^2} dx'\right) \frac{dx}{G(x)}$$
(A6)

which yields with the corresponding functions for $G(x)$ and $F(x)$:

$$P(x) \sim \frac{1}{x^{\left(1+\frac{A}{D}\right)}}$$
(A7)

**Figures**

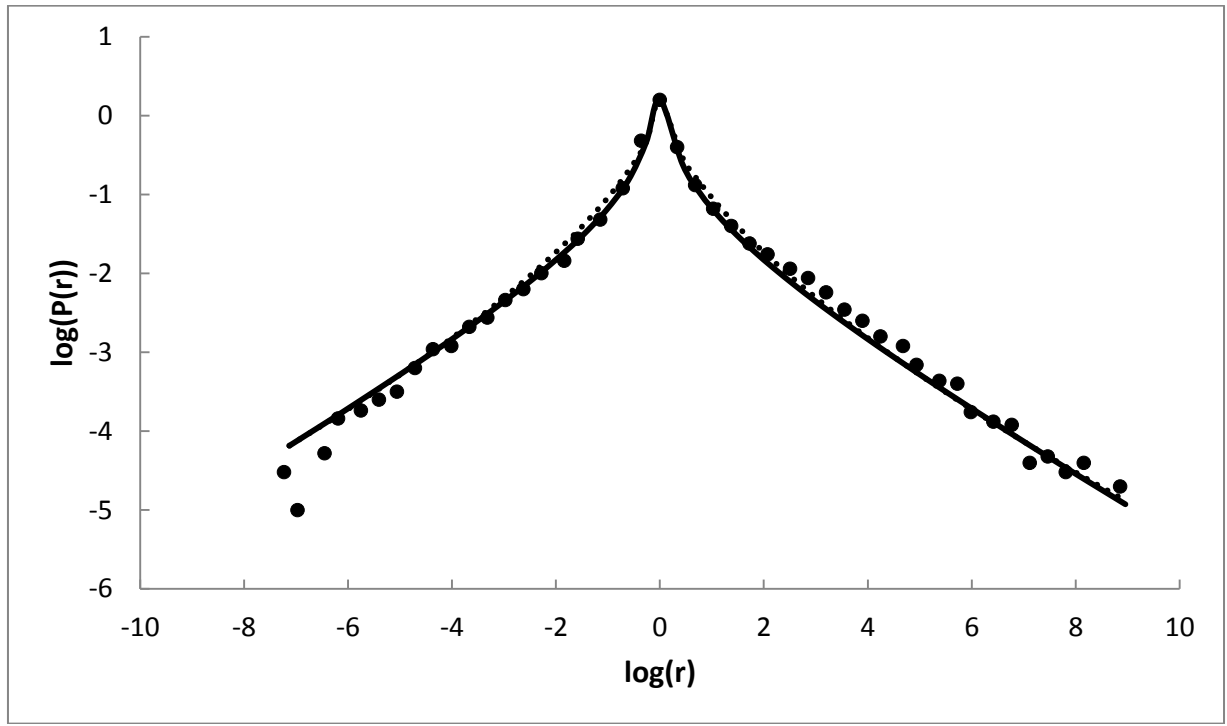

**Figure 1:** Growth rate distribution of pharmaceutical products (circles) [30], fitted by a Subbotin distribution (dotted line: $C_1=1.5$, $C_2=2.8$, $\beta=0.65$) and Eq. (87) (fat line: $C_r = 0.15$ and $\sigma_{rm}= 0.81$).

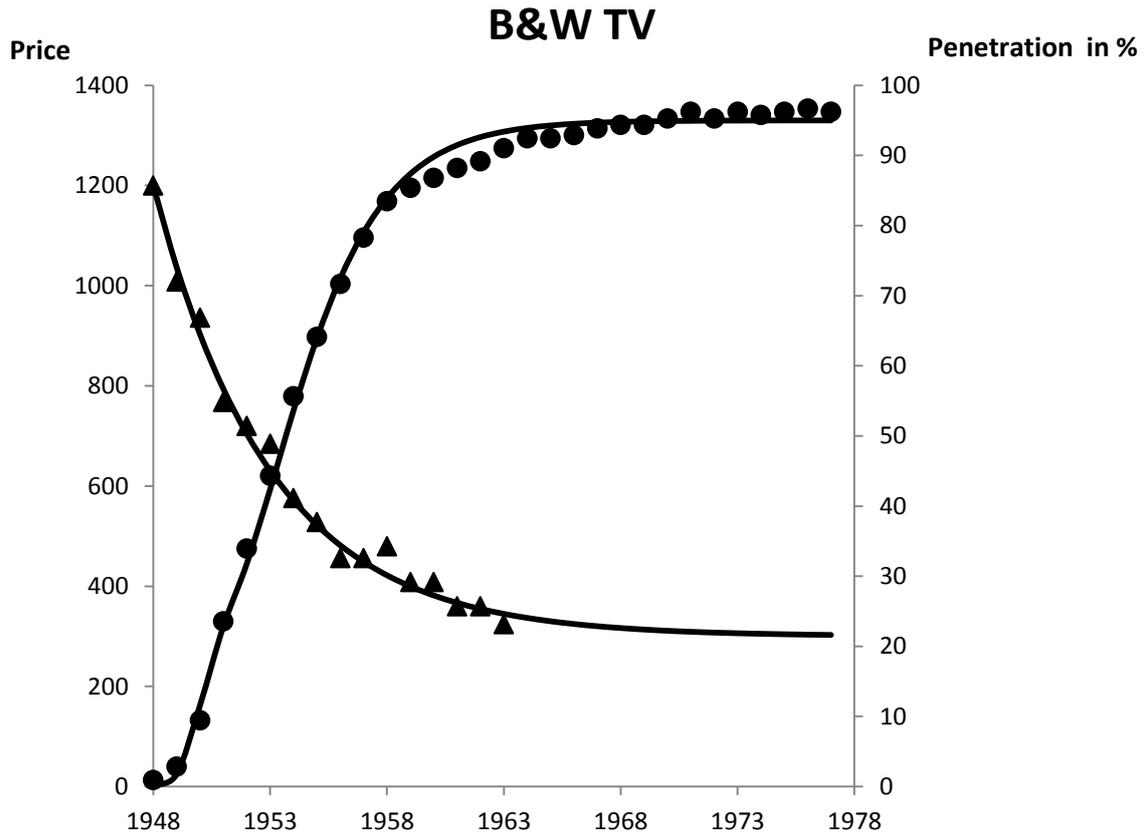

**Figure 2:** Evolution of the price in US $ (triangles) and the percentage market penetration (circles) of Black & White TV sets in the USA [1].

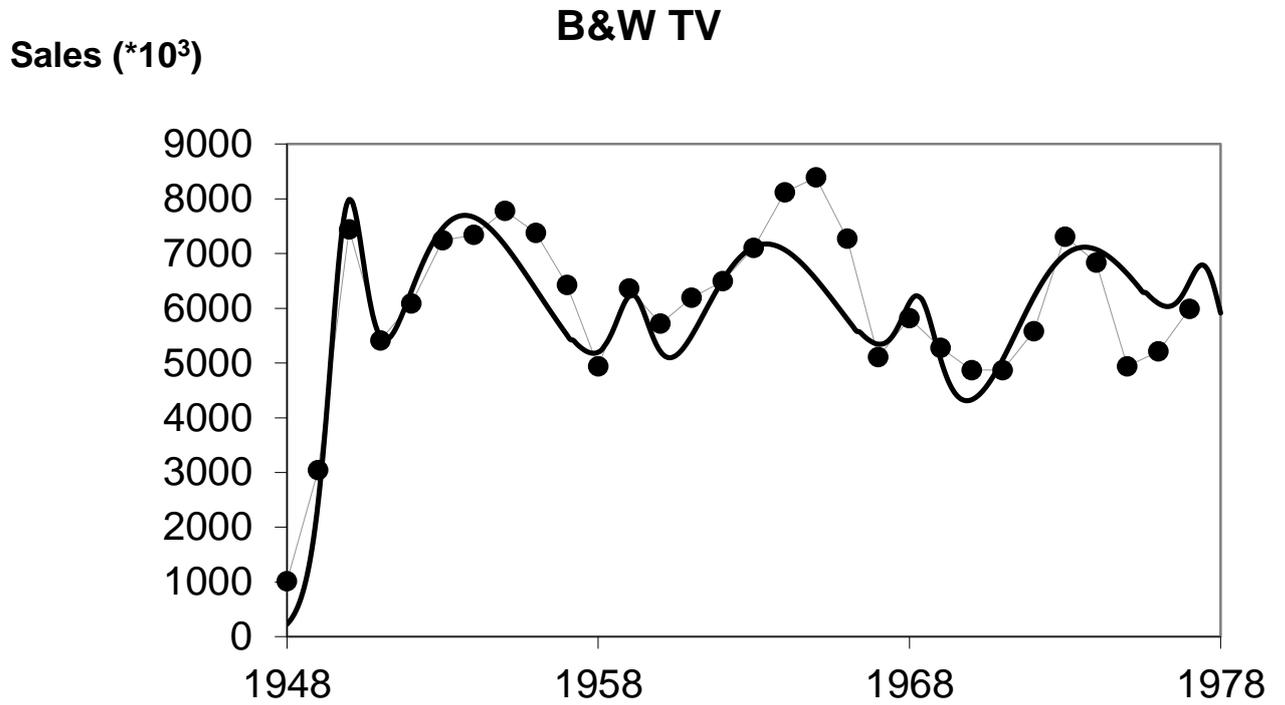

**Figure 3:** Evolution of the unit sales of Black & White TV sets in the USA. The fat line is a fit of the product life cycle [1].

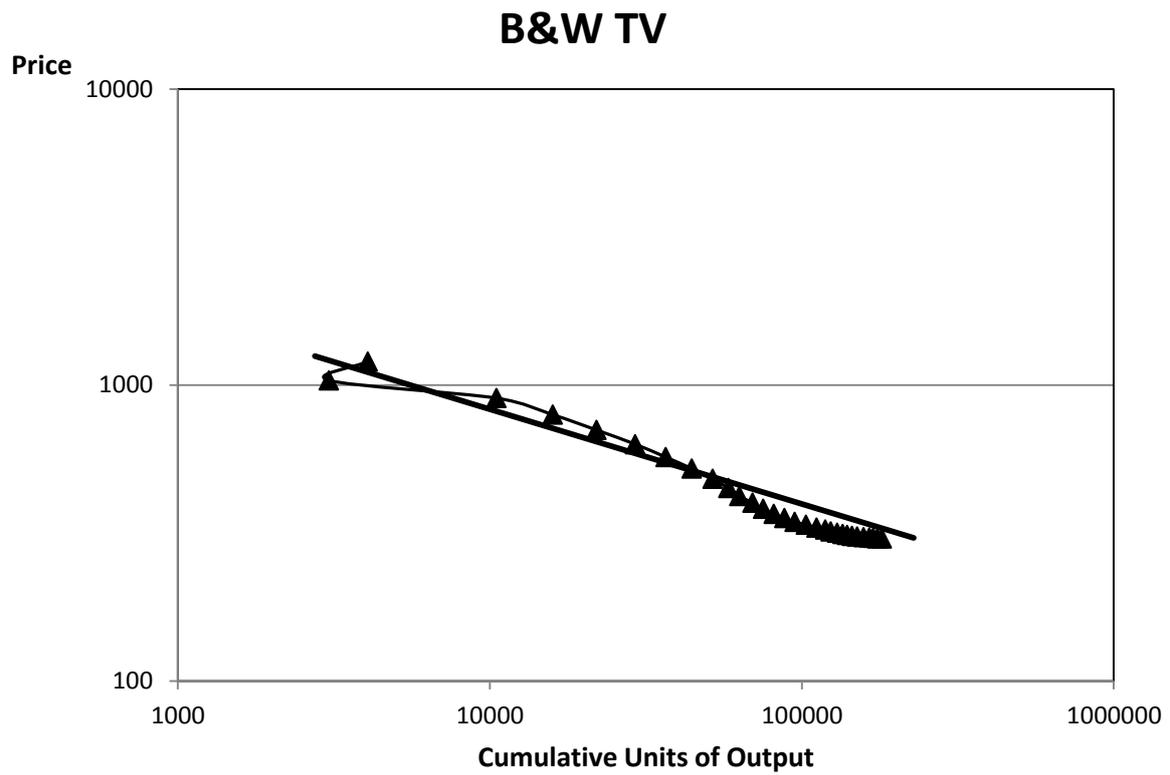

**Figure 4:** Experience curve of B&W TV sets in the USA. The triangles represent the price in US $ from Fig.2. The fat line expresses Henderson's law with 20% reduction of the price for every doubling of the cumulative output.

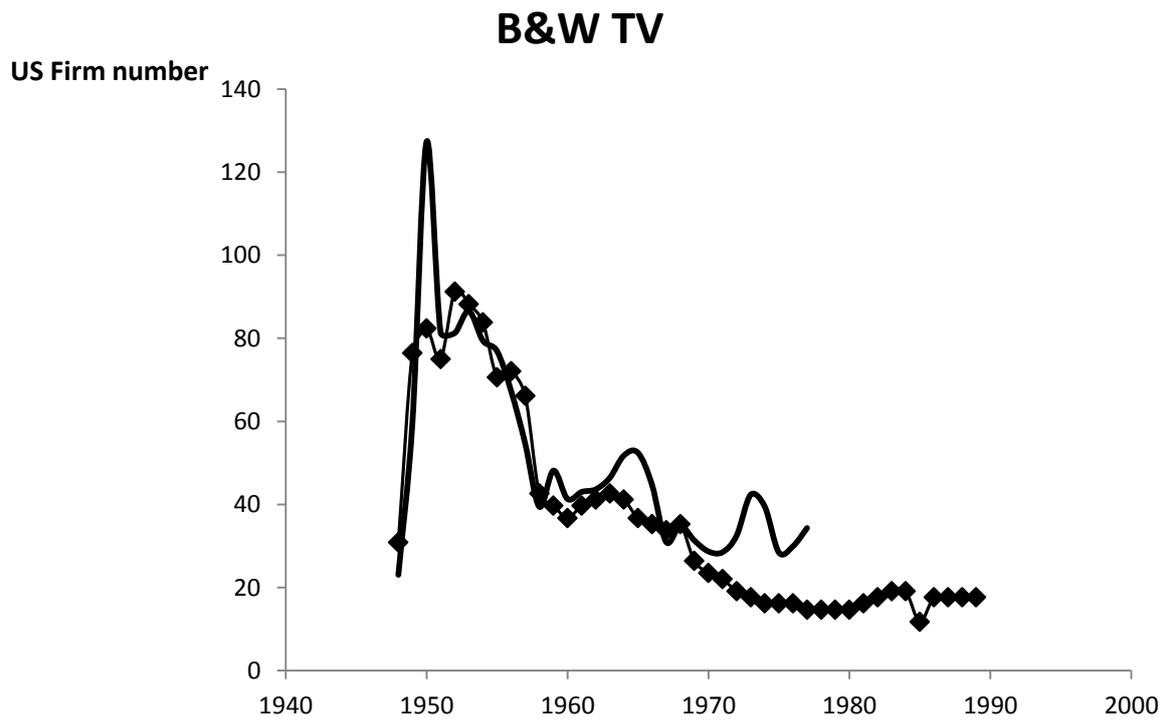

**Figure 5:** The market size of US Black & White TV sets [27]. The fat line represents Eq.(107) with $N_{f0}=0$ and $B \approx 1.8 \ 10^{-5}$ per US$.